\documentclass[prl,twocolumn,showpacs,amsmath,amssymb,superscriptaddress]{revtex4-2}
\usepackage{color}
\usepackage{amssymb}
\usepackage{amsthm}
\usepackage{graphicx}
\usepackage{epsfig}
\usepackage{subfigure}
\usepackage{amsmath,amssymb}
\usepackage[colorlinks=true,citecolor=blue,urlcolor=blue]{hyperref}
\usepackage{color}

\usepackage{bm}
\usepackage{tikz}
\usepackage{verbatim}

\usetikzlibrary{decorations}
\usetikzlibrary[decorations]
\usepgflibrary{decorations.pathmorphing} 
\usepgflibrary[decorations.pathmorphing] 
\usetikzlibrary{decorations.pathmorphing} 
\usetikzlibrary[decorations.pathmorphing] 

\begin{document}
\title{Non-Hermitian superfluid--Mott-insulator transition in the one-dimensional zigzag bosonic chains}
\author{Chengxi Li}
\affiliation{Beijing National Laboratory for Condensed Matter Physics,\\Institute of Physics, Chinese Academy of Sciences, Beijing 100190, China}
\affiliation{School of Physical Sciences, University of Chinese Academy of Sciences, Beijing 100049, China}
\author{Yubiao Wu}
\affiliation{Beijing National Laboratory for Condensed Matter Physics,\\Institute of Physics, Chinese Academy of Sciences, Beijing 100190, China}
\author{Wu-Ming Liu}
\email{wmliu@iphy.ac.cn}
\affiliation{Beijing National Laboratory for Condensed Matter Physics,\\Institute of Physics, Chinese Academy of Sciences, Beijing 100190, China}
\affiliation{School of Physical Sciences, University of Chinese Academy of Sciences, Beijing 100049, China}
\affiliation{Songshan Lake Materials Laboratory, Dongguan, Guangdong 523808, China}
\begin{abstract}
    We investigated the behavior of non-Hermitian bosonic gases with Hubbard interactions in the one-dimensional zigzag optical lattices through the calculation of dynamic response functions. 
    Our findings showed the existence of a non-Hermitian quantum phase transition that is dependent on the pseudo-Hermitian symmetry. 
    The system tends to exhibit a superfluid phase, when subjected to weak dissipation. 
    While under strong dissipation, the pseudo-Hermitian symmetry of the system is partially broken, leading to a transition towards a normal liquid phase. 
    As the dissipation increases beyond the critical threshold, the pseudo-Hermitian symmetry is completely broken, resulting in a Mott-insulator phase. 
    We propose an experimental setup using one-dimensional zigzag optical lattices containing two-electron atoms to realize this system. 
    Our research emphasizes the key role of non-Hermiticity in quantum phase transitions and offers a new theoretical framework as well as experimental methods for understanding the behavior of dissipative quantum systems, implicating significant  development of new quantum devices and technologies.
\end{abstract}


\maketitle

The superfluid--Mott-insulator phase transition in strongly correlated gases has been widely studied \cite{colussi2023lattice,fraxanet2022topological,basak2021strongly,meng2023atomic}.
In recent years, the zigzag bosonic chains has been mainly researched for strongly correlated quantum phase transitions and Majorana fermions \cite{kuhner2000one}. 
This system contains various Mott-insulator phases and gapless superfluid phases, forming a rich phase diagram \cite{vishveshwara2021z,han2015supersolid,liu2010simulating}. 
At the same time, the system also helps us to understand magnetic models and can be experimentally verified by implementing ultracold atoms in optical lattices \cite{shiba1972magnetic,chen2012kondo,zhang2015two,yan2017spin}. 
In the zigzag bosonic chains, the paired Bose-Hubbard model can be realized.
Unlike the Bose Hubbard model, the paired Hubbard model introduces a pairing contribution, which creates and annihilates boson pairs on adjacent lattice sites. 
This pairing contribution may lead to the appearance of $Z_2$ phase and topological properties \cite{vishveshwara2021z,zhao2021anomalous,zhao2021defective}. 
Therefore, this model provides rich resources for physical intuition and the connection with magnetic models \cite{peterson2008realizing}.

However, the researches on one-dimensional zigzag bosonic chains face a number of challenges in non-Hermitian systems. 
Firstly, the eigenvalues and eigenvectors of the system no longer have real properties, making the determination of phase transitions more complex \cite{brody2013biorthogonal, ohlsson2021density, roccati2022non}. 
Secondly, the properties of the superfluid and Mott-insulator phases of the system may also change. 
For example, the coherence of the superfluid phase may be lost, while the localization of the Mott-insulator phase may become more pronounced \cite{yamamoto2019theory,zhang2021probing}.
Finally, in terms of experiments, there are also challenges in the implementation and control of the one-dimensional zigzag bosonic chains in non-Hermitian systems.
For instance, different methods are required for the preparation and control of this model due to the non-real energy in non-Hermitian systems \cite{liang2022dynamic}. 
Additionally, dissipation in non-Hermitian systems may also have an impact on experimental results, necessitating more sophisticated experimental design and control.

In this Letter, we proposed and investigate the non-Hermitian one-dimensional zigzag bosonic chains in ultracold atoms. 
This work was inspired by a recent series of non-Hermitian transport discoveries \cite{lai2019photovoltaic,zhu2013spin,brody2013biorthogonal,ohlsson2021density, roccati2022non,sergi2013non,sergi2015time,ju2019non,wu2010interacting}. 
We utilized the theory of non-Hermitian linear response to calculate the Green's function of our system at zero temperature \cite{pan2020non,sticlet2022kubo}. 
By applying Kubo's formula, we analyzed the evolution and phase transition of the non-Hermitian system under external perturbation, and discovered that the non-Hermitian phase transition is accompanied by symmetry breaking.
We found that, the superfluid--Mott-insulator phase transition can be effectively controlled by dissipation.
These findings provide important insights for understanding phase transitions and critical phenomena in non-Hermitian many-body systems.



\begin{figure}[t]
    \centering
    \includegraphics[width=3in]{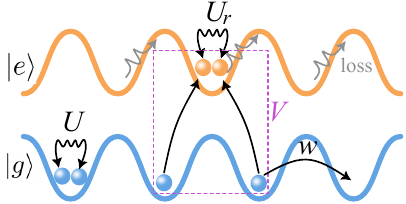}\\
    \caption{
    Implementation of non-Hermitian paired Bose-Hubbard model in the zigzag optical lattices. 
    A stable (dissipative) optical lattice is applied to the ground (excited) state $|g\rangle$ ($|e\rangle$). 
    $w$ is the hopping term for $|g\rangle$ chain and the inter-chain interaction $V$ can induced the pair hopping via an energy conserving process.
    The on-site interaction strength is $U(U_r)$ for $|g\rangle$ $(|e\rangle)$ chain. A running wave parallel to an optical lattice couples $|g\rangle$ to $|e\rangle$, which undergoes rapid on-site loss, giving rise to the non-hermitian pair hopping process. }
    \label{exp}
\end{figure}

\emph{Non-Hermitian zigzag bosonic chains.--}
Out experiment setup consist of two-level bosons with internal states $|g\rangle$ and $|e\rangle$ in the one-dimensional zigzag optical lattices. 
A typical lattice scheme is illustrated in Fig. \ref{exp}(a) for $^{174}$Yb, where the two levels are encoded in the atomic states $^1 S_0=|g\rangle$ and $^3 P_0=|e\rangle$.
Utilizing specific laser wavelength $\lambda_L=1120$ nm of optical lattice, for which the polarizations of $|g\rangle$ and $|e\rangle$ are opposite \cite{yi2008state,gerbier2010gauge}, 
we can generate internal-state-dependent potentials, and the potential minima for $|g\rangle$ chain locate in the middles of each of those for $|e\rangle$ chain.
The excited state is assumed to be unstable with a loss rate, and the chain loaded with $|e\rangle$ is treated as reservoir chain for concentrating on $|g\rangle$ chain of interest, and it can be described by the non-Hermitian paired Bose-Hubbard Hamiltonian
\begin{align}
        H&=-\sum_{\langle i,j\rangle}(w b_i^\dagger b_j + V b_i^\dagger b_j^\dagger b_{k,R} b_{l,R} + \text{h.c.}) \notag \\
        &+\sum_i \left[\frac{U}{2}n_i(n_i-1)-\mu_0 n_i\right].
        \label{eq-NHPBHM}
\end{align}
Here $b_i(b_i^\dag)$ is annihilation(creation) operator for atoms in  $|g\rangle$ chain at site $i$ 
with density operator $n_i = b_i^\dag b_i$, 
and $b_{k,R}$ and $b_{l,R}$ is that for atoms in reservoir chain at site $k$ and $l$.
$w$ is the hopping magnitude and $U$ is the on-site interaction strength for $|g\rangle$ chain.
$\mu_0$ is the tunable chemical potential offset between two chains and the chemical potential of reservoir chain is shifted to zero.
The pair hopping is introduced by an energy conserving process mediated by inter-chain interaction $V$.

In particular, the basic idea behind the atomic implementation of pairing terms can be engineered by applying a laser beam resonant to the energy difference, which offsets the interaction energy $U_r$ of two bosons on the reservoir chain and the chemical potential offset $\mu_0$ between the chains.
Thus, single-particle interchain tunneling is suppressed, and resonant pair tunneling to or from
a single site on the reservoir chain dominates, contributing a non-negligible quartic bosonic process and effectively giving a pairing term on $|g\rangle$ chain. 
By employing the mean-field approach to treat the interaction terms, 
we can obtain the order parameters $\Delta_{ij,1} =V\langle b_{ij,R}b_{ij,R}\rangle\approx \Delta_1$ and $\Delta_{ij,2}=V\langle b_{ij,R}^\dagger b_{ij,R}^\dagger\rangle \approx \Delta_2$.
Since the effective term depend on the expectation value of the pair annihilation or creation process on reservoir chain with a on-site loss rate, the pairing parameters will be imbalanced,
and we define them as unequal real parameters which can be obtained by $U(1)$ transformation.

In the strongly interacting limit, the Hilbert space of our system can be restricted to the number-basis states 
$|n_0\rangle$ or $|n_0+1\rangle$ at each sites.
Such a hard-core bosonic model can be transformed into a magnetic model using the operator representation $s=1/2$, $|n_0+1\rangle=|\uparrow\rangle$,$|n_0\rangle=|\downarrow\rangle$ 
by defining $b_i^\dagger b_j\rightarrow (n_0+1)s_i^+ s_j^-,n\rightarrow n_0+s^z+1/2$.
Using degenerate perturbation theory to second order of $w/U$, the transverse field XY model can be derived, with complex anisotropic spin exchange integrals $J_x$ and $J_y$,  and Zeeman magnetic field $h$ \cite{vishveshwara2021z, supplement}.
Then it can be transformed into the quasifermionic Effective Kitaev model (EKM) by a Jordan-Wigner transformation $s_i^+=c_i^\dagger e^{-i\pi\sum_{j<i}c_j^\dagger c_j}$ \cite{supplement} in the $t/U\to0$ limit,
i.e., $H=\sum_{\langle i,j\rangle}-t c_i^\dagger c_{j}+h.c.+\Delta c_{i}^\dagger c_j^\dagger+\gamma c_i c_{j}+\sum_{i}\mu c_{i}^\dagger c_{i}$. 
Here $t=(n_0+1)w$ is the nearest-neighbor hopping amplitude, $\Delta=(n_0+1)\Delta_1$ and $\gamma=(n_0+1)\Delta_2$ denotes the strength of pair parameters between the nearest-neighbor sites,
and $\mu$ is the on-site renormalized chemical potential.
Although the superfluid order parameters in this case are complex, they can always be transformed into real through a unitary transformation \cite{supplement}.
In the following, we calculate the linear response of EKM to investigate the physical properties of our system near the Mott-insulator regime.

\begin{figure}[t]
    \centering
    \includegraphics[width=3.4in]{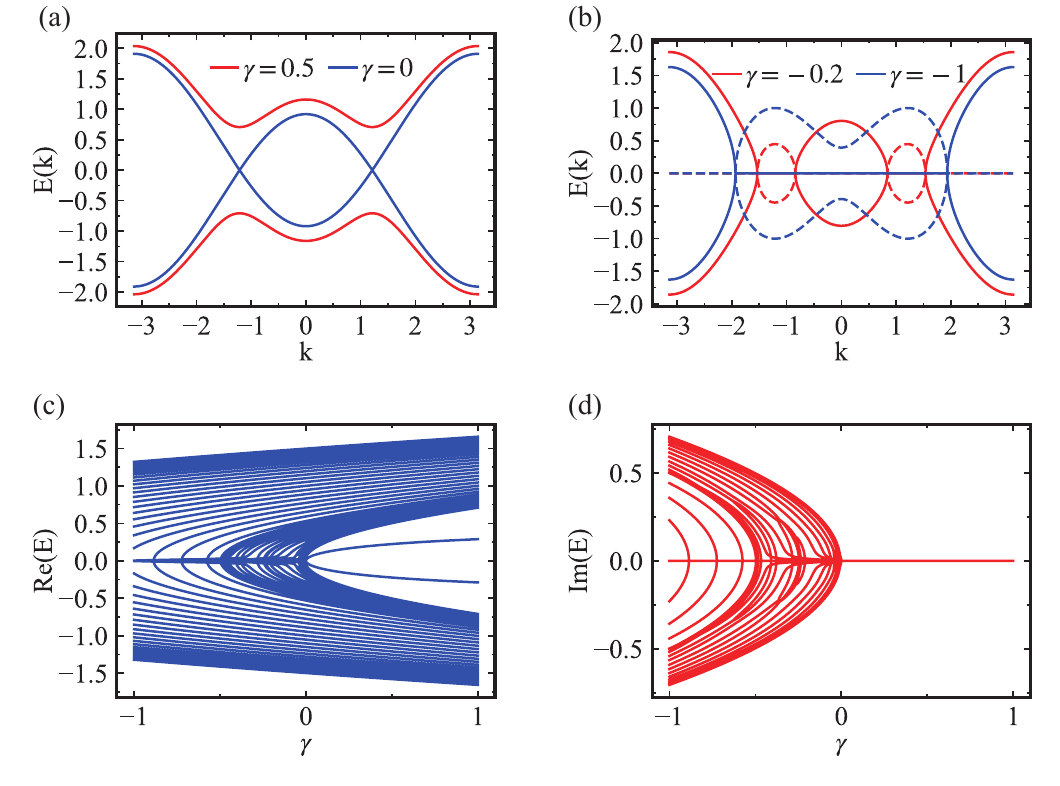}
    \caption{   
    (a)--(b) Energy spectrum diagram in k-space with $\gamma>0$ in (a) and $\gamma<0$ in (b).
    The full line denotes the real part of energy and the dashed line denotes imaginary part of energy.  
    The fully real energy spectrums in (a) implies there is a pseudo-Hermitian symmetry when $\gamma>0$, while that pseudo-Hermitian symmetry is broken when $\gamma<0$ in (b).
    Other parameters are set as $t=1,\Delta=0.5,\mu=0.7$.
    (c)-(d) Energy spectrum diagram in real space as a function of $\gamma$. 
    As $\gamma$ decreases, purely imaginary eigenvalues appear, indicating a transition of the system from complete pseudo-Hermitian symmetry to partial pseudo-Hermitian symmetry breaking. } 
    \label{energy spectrum}
\end{figure}

\emph{Pseudo-Hermitian symmetry.--}
Under periodic boundary conditions, the PT symmetry breaking Hamiltonian becomes $H(k)=\xi(k)\sigma_z/2+(\Delta+\gamma)\sigma_x/2+i(\Delta-\gamma)\sigma_y/2$ with $\xi(k)=\mu-2t\cos k$ and $\sigma_{x,y,z}$ are Pauli matrices on unit cell basis.
However, in the regime where the energy spectrum is real, the Hamiltonian preserves pseudo-Hermitian symmetry \cite{cherbal2016Pseudo}.
The eigenvectors $|u_+\rangle$ and $|u_-\rangle$ correspond to the two bands $E_{\pm}(k)=\pm\epsilon(k)$, where $\epsilon(k)=\sqrt{\xi(k)^2/4+\Delta\gamma}$ [see Fig. \ref{energy spectrum}].
The $\eta$-pseudo-Hermitian symmetry of the system, where $\eta=\eta_+$, has been extensively studied in previous works \cite{cherbal2016Pseudo,ohlsson2020transition,ohlsson2021density,lein2019krein,wang2020Pseudo}.
For energy values satisfying $\epsilon(k)^2>0$, the metric operator $\eta_+$ is defined by \cite{cherbal2016Pseudo} $\eta_+^{-1}=|u_+\rangle\langle u_+|+|u_-\rangle\langle u_-|$, where $|u_+\rangle$ and $\langle u_-|$ denote the eigenvectors corresponding to the two bands $E_\pm(k)=\pm\epsilon(k)$.
Upon substituting these eigenvectors into the aforementioned formula and simplifying, we obtain the explicit expression for the metric operator $\eta_+$, whose inverse is given as
\begin{equation}
    \eta_+^{-1}=\begin{pmatrix}
        \frac{(\xi^2(k)+2\Delta^2+2\Delta\gamma)}{\xi^2(k)+(\Delta+\gamma)^2}&\frac{\xi(k)(\gamma-\Delta)}{\xi^2(k)+(\Delta+\gamma)^2}\\
        \frac{\xi(k)(\gamma-\Delta)}{\xi^2(k)+(\Delta+\gamma)^2}&\frac{\xi^2(k)+2\gamma^2+2\gamma\Delta}{\xi^2(k)+(\Delta+\gamma)^2}
    \end{pmatrix}     \,,
\end{equation}
It satisfies the relation $\eta_+ H\eta_+^{-1}=H^\dagger$, and its determinant can be expressed as $\rm{det}(\eta)=\frac{\xi^2(k)+(\Delta+\gamma)^2}{\xi^2(k)+4\Delta\gamma}$.
Since the energy spectrum of the system is real, i.e., when the square of energy satisfies $\xi^2(k)+4\Delta\gamma>0$, it follows $\det(\eta)>0$. 
This property allows for the well-defined nature of the $\eta_+$-inner product \cite{ohlsson2020transition}. 
Specifically, the $\eta_+$-inner product can be expressed as $\langle \psi|\phi\rangle_+=\langle\psi|\eta_+|\phi\rangle$, where the set of states $\{|\phi\rangle\}$ form a Hilbert space $\mathcal{H}$ equipped with the metric $\eta_+$ \cite{choutri2017Pseudo}.

The Hamiltonian in the presence of a gap exhibits a pseudo-Hermitian symmetric phase for $\Delta\gamma>0$ due to the invariance of eigenvectors under the pseudo-Hermitian symmetry operator $\eta H^\dagger(k)\eta^{-1}=H(k)$ and the existence of two real energy bands in the spectrum \cite{mostafazadeh2002Pseudo}.
The spectrum becomes gapless with a linear dispersion $\epsilon(k)=t\sin k_0|k\pm k_0|$ at $\gamma\Delta=0$, where $k_0$ denotes an energy gapless point $(-k_\mathcal{EP},k_\mathcal{EP})$.
An exceptional point (EP) is marked at $k=k_0$ where biorthogonal Hilbert spaces lose their completeness. 
In the region of real eigenvalues, the system is in a pseudo-Hermitian symmetry phase if $\cos k<(\mu-\sqrt{-4\Delta\gamma})/2$ or $\cos k>(\mu+\sqrt{-4\Delta \gamma})/2t$, while it is in a pseudo-Hermitian symmetry broken phase  with conjugate pairs of imaginary eigenvalues if $(u-\sqrt{-4\Delta\gamma})/2t<\cos k<(\mu+\sqrt{-4\Delta\gamma})/2t$ for $\Delta\gamma<0$.
The regions with real spectra are separated by EPs occurring at $\epsilon(k_0)=0$ [Fig. \ref{energy spectrum}].

\emph{Quantum transport.--}
On the non-Hermitian physics, the response function does not possess time translation invariance owing to the varying density matrix $\rho(t)$ at various moments, which may exhibit a non-unitary evolution even in the absence of external perturbations.
To address this issue, Sticlet et al. \cite{sticlet2022kubo} and Pan et al. \cite{pan2020non} have introduced a generalized response function for non-Hermitian systems, given by $\chi_{A,B}(t,t')=-i\theta(t-t')tr\{[A(\tau),B]-\langle A(t)\rangle_0[e^{iH_0^\dagger\tau}e^{-iH_0},B]\frac{\rho_0(t')}{tr{\rho_0(t)}}\}$.
Herein, $[A,B]$ denotes the so-called generalized commutator defined as $[A,B]=AB-B^\dagger A$.
$\tau=t-t'$ represents the time interval between the initial and final states, and $\rho_0$ signifies the density matrix of the unperturbed system Hamiltonian $H_0$, which evolves with time. 
The formulation is obtained using the right eigenvectors of $H_0$.

In general, non-Hermitian dynamics may exhibit non-unitary evolution, and the system's density matrix is not necessarily fixed at its initial value ($\rho(t)\neq\rho_0$).
Nonetheless, certain physical systems possess pseudo-Hermitian symmetry, which causes the non-Hermitian density matrix to remain unchanged at its initial value \cite{supplement}.
For instance, in the pseudo symmetric phase of our system, the density matrix becomes independent of time.

In the presence of an external potential $V(t)$, the pseudo-Hermitian model exhibits a real eigenspectrum. 
The influence of an external field on a neutral atomic system can be characterized by the particle current density operator $j = -\delta H/\delta V$.
Despite the non-Hermitian nature of the Hamiltonian, the particle current operator remains Hermitian.
In this scenario, which can be regarded as a time-independent perturbation, the response function is given by:
\begin{align}
        \chi(\tau)=&\sum_{k\in PS}i\theta(\tau)\langle[j(k,\tau),j(k,0)]\rangle_0    \notag \\
        &-\langle j(k,0)\rangle_0\langle[e^{iH_0^\dagger\tau}e^{-iH_0\tau},j(k,0)]\rangle_0
		\,,   \label{response}
\end{align}
where the current operator $j$ is defined as $j=-1/2\sum_{k}\partial \xi(k)/\partial k \sigma_z$.
Herein, the sum over $k$ can only be evaluated in the real spectrum, and the thermal expectation value is defined by $\langle j\rangle_0=tr(\rho_0 j)/tr(\rho_0)$.
The response function assumes a non-zero value solely in the pseudo-Hermitian symmetry phase, while remaining zero in the broken pseudo-Hermitian symmetry phase.
This result follows from an exact cancellation between the generalized commutator contribution to the response, $[j(k,\tau),j(k,0)]$, and the norm corrections \cite{supplement}.
At zero temperature, the system exhibits a half-filled Majorana zero mode \cite{zhao2021defective}.

For generic real pairing order parameters, both the generalized commutator and the norm correction terms contribute to $\chi(k,\tau)$, which can be expressed as \cite{supplement}:
\begin{align}
        &\chi(k,\tau)=\theta(\tau)\sin^2 k\sin 2\epsilon(k)\tau A[\epsilon(k),\xi(k)] \,,  \notag  \\
        &A[\epsilon(k),\xi(k)]=\frac{128 t^2\gamma^2(2\epsilon-\xi)^2}{[4\gamma^2+(2\epsilon(k)-\xi(k))^2]^2} \,,
\end{align}
where $\epsilon(k)=\sqrt{\xi^2(k)/4+\Delta\gamma}$ represents the absolute value of the energy eigenvalues.
Notably, the Hamiltonian $H(\gamma=-\Delta)=\xi(k)/2\sigma_z+i\Delta\sigma_y$ implies that $[j(k),H]=0$, thereby rendering the current operator time-independent.
As a result, the first contribution to $\chi(k,\tau)$ in Eq. (\ref{response}), involving the commutator $[j(k,\tau),j(k)]=0$, is discarded.
However, the second contribution cannot be ignored.
This effect is entirely non-Hermitian, a feature absent from Hermitian systems.

\begin{figure}[t]
    \centering
    \includegraphics[width=3.4in]{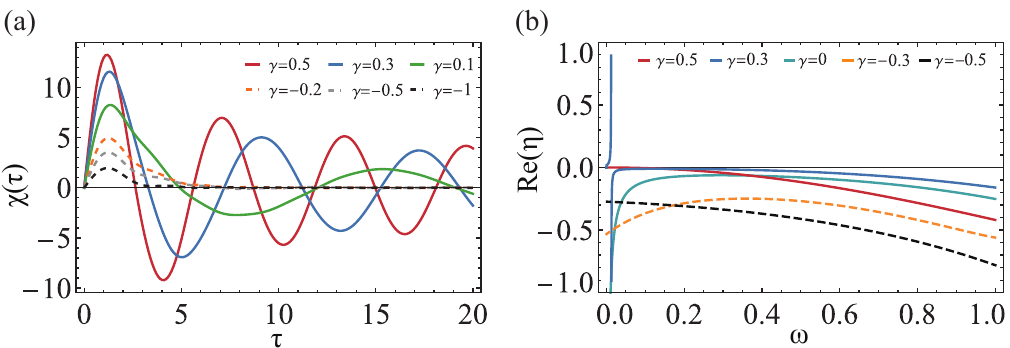}
    \caption{(a) The spatio-temporal evolution of the response function $\chi(\tau)$ is shown, where $\tau$ represents time and $\gamma$ denotes the dissipation parameter. 
        (b) The numerical solution of the viscosity of the one-dimensional non-Hermitian zigzag bosonic chains in the frequency domain is presented for varying levels of dissipation strength. 
        In the case of weak dissipation $0<\gamma<\Delta$, the viscosity vanishes in the low frequency limit. 
        In contrast, for strong dissipation $\gamma<0$, the viscosity in the low frequency regime remains non-zero.}
    \label{rf_time}
  \end{figure}  

\emph{Superfluid viscosity.--}
The time-dependent response function $\chi(\tau)$ is obtained by summation over all momentum states [Fig. \ref{rf_time}(a)].
Our findings show that in the pseudo-Hermitian symmetry phase, the dissipation parameter $\gamma$ is negative and $\chi(\tau)$ exhibits damped oscillation with slow decay. 
While in the pseudo-Hermitian symmetry broken phase, the $\gamma$ is positive and the response function decays exponentially. 
The response in frequency space, $\chi(\omega)$, is obtained by Fourier transforming Eq. (\ref{response}) and summing over momenta, i.e., $\chi(\omega)=\int_{k\in PS}\chi(k,\omega)$, where $\eta=0^+$. 
The real part of viscosity is given by the imaginary part of the response function, i.e., $1/\eta'(\omega)=\chi''(\omega)/\omega$, where $\chi''(\omega)=[\chi(\omega)-\chi(-\omega)]/2$ \cite{geier2021non}. 
After lengthy calculation, the real part function of viscosity can be expressed in \cite{supplement}.
The viscosity curves of the EKM to an external potential are shown in Fig. \ref{rf_time}(b).

The effects of dissipation on the viscosity response can be explained by the interplay between the superfluid and dissipation \cite{leggett1980cooper, leggett1980diatomic, nozieres1985bose}.
In the limit of weak dissipation ($0<\gamma<\Delta$), superfluids dominate, and the system retains its energy gap.
When the frequency of the external potential exceeds the energy gap, Cooper pairs are destroyed, causing  superfluids unstable under high-frequency external potential.
In the case of strong dissipation ($\gamma>0$), the system is governed by the continuous quantum Zeno effect (QZE) \cite{syassen2008strong,mark2012preparation,barontini2013controlling,zhu2014suppressing,tomita2017observation}, suppressing neighboring tunneling, leading to particle localization, and coherence of Cooper pairs is suppressed by dissipation, ultimately destroying superfluid.

\begin{figure}[t]
    \centering
    \includegraphics[width=3.4in]{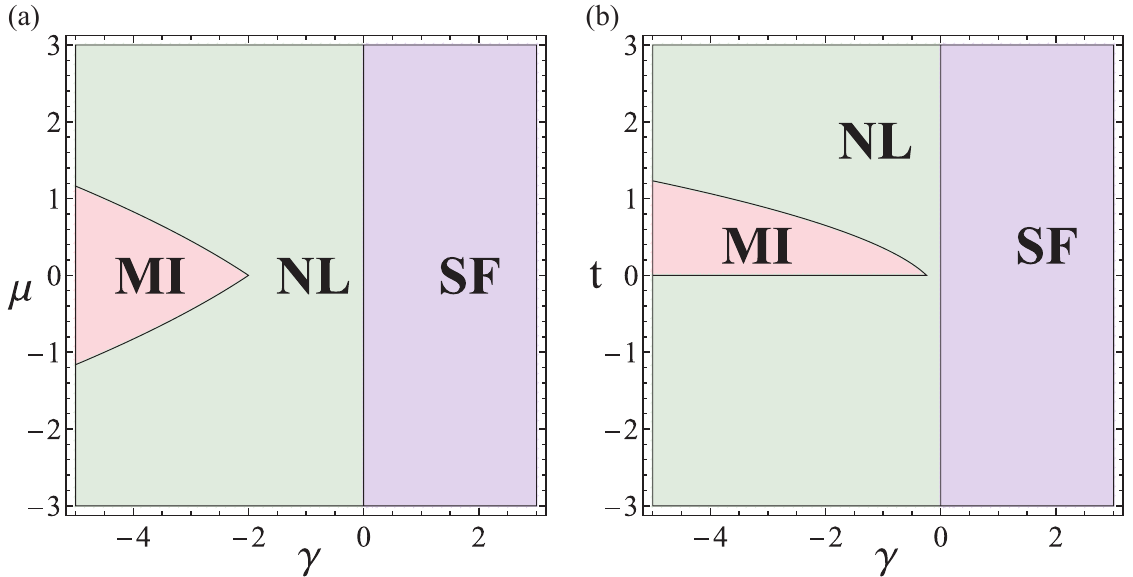}
    \caption{ (a) Phase diagram of the system parameterized by dissipation strength $\gamma$ and chemical potential $\mu$.
    As $\gamma$ increases below zero, the system undergoes a transition from the Mott-insulator (MI) phase to the normal liquid (NL) phase. 
    When $\gamma$ exceeds zero, the system undergoes a transition from the normal liquid phase to the superfluid phase (SF).
    (b) Phase diagram of the system parameterized by dissipation strength $\gamma$ and tunneling $t$.
    When $\gamma>0$ superfluid phase appears, while $\gamma<0$ normal liquid phase appears. 
    With $\gamma$ continues decreasing, the positive pseudo-Hermitian symmetry is completely broken, and the system turns into Mott-insulator phase.} 
    \label{phase_diagram}
\end{figure}

At the critical point $\gamma=0$, the system viscosity is divergence, indicating a transition from a superfluid phase to a normal liquid phase.
In this case, the energy spectrum becomes a linear dispersion relation $\epsilon(k)=|\xi(k)|/2\simeq v|k-k_0|$ (see Fig. \ref{energy spectrum}) near the exceptional points (EPs) $k_0$, where $v=\partial\epsilon(k)/\partial k=t\sin k_0 \mathrm{sgn}(k-k_0)$ represents the group velocity.
Then Hamiltonian can be expressed as
\begin{equation}
    H(k)=\begin{pmatrix}
        v|k-k_0|&\Delta\\
        &-v|k-k_0|
    \end{pmatrix}  \,.
\end{equation}

\noindent
There are only two EPs. 
At EPs, the Hamiltonian eigenvectors no longer constitute a complete Hilbert space, and there is only one eigenvector, $|+\rangle=(1,0)^T$. 
The particle current density is expressed as $j=v\langle+|\sigma_z|-\rangle=t\sin k_0\mathrm{sgn}(k-k_0)$. 
Thus, in the case of $|k|>|k_\mathcal{EP}|$, the particle current density is $j=t\sin k_\mathcal{EP}$, while in the case of $|k|<|k_\mathcal{EP}|$, the particle current density is $j=t\sin k_\mathcal{EP}$.
In momentum space, currents in different regions cancel each other out.

In the non-Hermitian zigzag bosonic chains, when the dissipation strength is sufficiently large, the quasi-fermion are bound to the protocell due to the continuous QZE. 
Based on our findings, we depict phase diagrams of our system in Fig. \ref{phase_diagram}. 
The purple region denotes the system remaining in the superfluid phase under weak dissipation conditions. 
The green region represents the superfluid phase being destroyed and replaced by a normal liquid phase under strong dissipation. 
In the red region, the system has no real energy spectrum and no pseudo-hermitian symmetry, and the system enters into the Mott-insulator phase.

\emph{Experimental realization.--}
Retuning to a realistic experiment with concrete parameters, we choose lattice trap potential $V_0=5E_R$ with $E_R=h^2/(2m\lambda_L^2)\approx 0.914$ kHz being the recoil energy as the energy unit.
The intrachain hopping amplitude can be determined as $t\approx 0.0192E_R \approx 17.55$ Hz.
Using the standard technique of Feshbach resonance, we can tune the bare interaction strength $g=-2.0E_R$,
which leads to the Hubbard interaction strength $U\approx -8.14E_R \approx 7.44$ kHz, satisfying the limit of $U \gg t$ \cite{greiner2002quantum,anderlini2007controlled,sugawa2011interaction}.
On the other hand, the interaction strength $U_r$ for reservoir chain can also be artificially tuned.
We can choose a fixed chemical potential offset $\mu_0$ between the chains and apply a running laser with frequency matching the energy difference $U_r-\mu_0$.
This interaction-induced resonant tunneling has been implemented in experiments to investigate interacting dipoles \cite{simon2011quantum,bissbort2012effective}.
At last, the imbalanced pairing parameters are induced by applying an additional laser with wavelength $\lambda=1285$ nm to couple $^3P_0$ to $^1P_1$ resonantly, and the complex part of pairing term can be tuned by the loss rate of excited state $|e\rangle$ \cite{xu2017weyl,gong2018topological,lapp2019engineering}.
The dynamical displacement of atoms can be measured by single-site resolved quantum gas microscopy \cite{miranda2015site}, and the transport response can be revealed by  spectroscopic probes \cite{greiner2002quantum,campbell2006imaging}.

In summary, we have investigated the non-Hermitian superfluid--Mott-insulator transition of the zigzag bosonic chains with strong Hubbard interaction.
Our work has revealed that such the quantum phase transition can be described by the transport behavior of the one-dimensional non-Hermitian zigzag bosonic chains under an alternating external potential.
Our non-Hermitian systems are constructed from dissipative open systems, and the dissipation strength plays a crucial role in determining the non-Hermitian response function's impact.
Specifically, in the weak dissipation limit, the system retains its energy gap, the system approaching the superfluid phase. 
Whereas in the strong dissipation limit, the QZE suppressing neighboring tunneling, the system tends toward the normal liquid phase.
As the dissipation increases further, Cooper pairs of quasi-fermions in the system decomposed, leading to a normal liquid-Mott insulator phase transition.
The calculation of non-Hermitian Green function has revealed that the viscosity of our system is contingent upon the emergence of pseudo-symmetries, which is a unconventional quantum phase transition characteristic of the non-Hermitian zigzag bosonic chains.
Notably, these phase transitions are accompanied by the appearance and disappearance of EPs.
These characteristics can be experimentally tested, and future work can further extend the one-dimensional zigzag bosonic chains to higher dimension.


\emph{Acknowledgements.--} 
This work was supported by National Key R\&D Program of China under grants No. 2021YFA1400900, 2021YFA0718300, 2021YFA1400243, NSFC under grants Nos.  61835013, 12174461, 12234012, Space Application System of China Manned Space Program.

\bibliographystyle{apsrev4-2}
\bibliography{ref}

\onecolumngrid
\section{Supplemental}
\section{Non-Hermitian paired-Bose-Hubbard model}

Considering the two-body loss in the optical lattice, the non-Hermitian paired-Bose-Hubbard model (NHHM) Hamiltonian obtained can be written as \cite{yamamoto2019theory,daley2014quantum} 
\begin{equation}
    H=-\sum_{i,j}(wb_i^\dagger b_j+h.c.+\Delta_1 b_i^\dagger b_j^\dagger+\Delta_2 b_jb_i)+\sum_i\left[\frac{U}{2}n_i(n_i-1)-\mu n_i\right],
\end{equation}
where the $U$ is repulsion interaction, $w$ is the hopping amplitude from site $j$ to $i$, $\mu$ is chemical potential and it's worth noting that $t,U,\Delta_{1,2}\in \mathbb{R}$. 
In the situation of strong coupling limit of the Hubbard model $(U\geq w)$, this is then a model of “hard-core” bosons with an infinite on-site repulsion energy.
The only states with a finite energy are those with $|n_0\rangle$ or $|n_0+1\rangle$ on every site of the lattice.
There are only two Mott insulators with $n=n_0$ or $n=n_0+1$ are permitted.
The hard-core paired Bose-Hubbard model can also be written as a magnet model ($s=1/2$) by taking transformation $|n_0+1\rangle=|\uparrow\rangle,|n_0\rangle=|\downarrow\rangle,b_i^\dagger b_j=(n_0+1)s_i^+ s_j^-$ and $n=n_0+1/2+s_z$ \cite{sachdev1999quantum}.
Thus the transverse field non-Hermitian XY model can be derivated by applying degenerate perturbation theory to the second order of $\tilde{t}/U$ \cite{vishveshwara2021z}[supplement] 
\begin{equation}
    H=\sum_{\langle i,j\rangle}\left[J_x s_i^x s_j^x+J_ys_i^y s_j^y-iJ_{xy}(s_i^xs_j^y+s_i^ys_j^x)\right]-h\sum_{i}s_i^z,
\end{equation}
where $J_x=-(n_0+1)(2w+\Delta_1+\Delta_2)$, $J_y=-(n_0+1)(2w-\Delta_1-\Delta_2)$ is anisotropic spin exchange integral, $J_{xy}=2(\Delta_1-\Delta_2)$ is non-Hermitian term and $h=\mu-U n_0$ is Zeeman magnetic field. 

The non-Hermitian XY model possesses pseudo-Hermitian symmetry, $\eta=\prod_{i}(-1)^{n_i}$, $H^\dagger=\eta H\eta^{-1}$, which ensures the possibility of a purely real energy spectrum.
The transverse field spin chain can be reformed to fermionic Kitaev model by Jordan-Wigner transformation $s_i^+=c_i^\dagger \exp(i\pi\sum_{j<i}c_j^\dagger c_j)$ and $s_i^-=\exp(-i\pi\sum_{j<i}c_j^\dagger c_j)c_i$. 
In this way, the non-Hermi Hubbard interaction is transformed into the Kitaev model in the $t/U\to 0$ limit. 
By calculating the linear response of the Kitaev model, the physical properties of the NHHM near the mott insulating phase can be obtained by 
\begin{equation}
    H=\sum_{\langle i,j\rangle}\left(-t c_i^\dagger c_{j}+h.c.+\Delta c_{i}^\dagger c_j^\dagger+\gamma c_i c_{j}\right)+\sum_{i}\mu c_{i}^\dagger c_{i},
\end{equation} 
where the $t=(n_0+1)w$ is hopping amplitude between the neighborest site, the $\Delta=(n_0+1)\Delta_1$ and $\gamma=(n_0+1)\Delta_2$ denotes the strength of pair parameters between the nearest-neighbor sites and the $\mu$ is on-site chemical potential. 

\section{Pseudo-symmetry system}
Since H is non-Hermitic, it has a biorthogonal basis $|u\rangle,|v\rangle$, which satisfies
\begin{equation}
    \begin{split}
        H|u_i\rangle&=E_i|u_i\rangle,\quad \langle u_i|H^\dagger=E_i^*\langle u_i|,\\
        H|v_i\rangle&=R_i|v_i\rangle,\quad \langle v_i|H^\dagger=R_i^*\langle v_i|,
    \end{split}
\end{equation}
\begin{equation}
    \langle v_i|u_j\rangle=\delta_{ij},
\end{equation}
The metric operator $\eta=\sum_i|v_i\rangle\langle v_i|$ and $\eta^{-1}=\sum_i|u_i\rangle\langle u_i|$ related $|u_i\rangle$ and $|v_i\rangle$
\begin{equation}
    \begin{split}
        |v_i\rangle=\eta|u_i\rangle,\quad |u_i\rangle=\eta^{-1}|v_i\rangle,
    \end{split}
\end{equation}

Using this relation, we can define the eta inner product between the Hilbert space $\mathcal{H}$ corresponding to the eigenvector $|u_i\rangle$ and the Hilbert space $\mathcal{H}^*$ corresponding to the eigenvector $\langle v_i|$
\begin{equation}
    \langle v_i|u_j\rangle=\langle u_i|\eta|u_j\rangle=\langle u_i|u_j\rangle_\eta=\delta_{ij},
\end{equation}
where $\langle \cdot|\cdot\rangle_\eta$ is called $\eta-$product. 

It has been proven in \cite{mostafazadeh2002pseudo,mostafazadeh2002pseudo2,mostafazadeh2002pseudo3} that the necessary and sufficient conditions for a non-Hermitian but diagonalizable Hamiltonian to have real eigenvalues is the existence of a linear positive-definite operator $\eta(\det\eta>0)$ such that $\eta H\eta^{-1}=H^\dagger$ is fulfilled.

The quasi-fermion EKM in Lattice space is written as 
\begin{equation}
    \begin{split}
        H=\sum_{\langle i,j\rangle}-t c_i^\dagger c_{j}+h.c.+\Delta c_{i}^\dagger c_j^\dagger+\gamma c_i c_{j}+\sum_{i}\mu c_{i}^\dagger c_{i}.
    \end{split}
\end{equation}
Taking Fourier transformation $c_n=1/\sqrt{N}\sum_{k}e^{ik\cdot R_n} c_k$ the Hamiltionian in k-space described by
\begin{equation}
    H=\sum_{k}\begin{pmatrix}
        c_k^\dagger & c_{-k}
    \end{pmatrix} \begin{pmatrix}
        \xi(k)/2 & \Delta\\
        \gamma & -\xi(k)/2
    \end{pmatrix}\begin{pmatrix}
        c_k\\
        c_{-k}^\dagger
    \end{pmatrix},
\end{equation}
where the $\xi(k)=\mu-2t\cos k$ and $H(k)=\xi(k)\sigma_z/2+(\Delta+\gamma)\sigma_x/2+i(\Delta-\gamma)\sigma_y/2$. The lattice constant we defined is unit and the lattice constant $a$ and Plank constant $\hbar$ we set unit$(a=\hbar=1)$.
The eigen value and the eigen vector of H can be solved respectively to obtain.
\begin{equation}\label{eigen}
    E_\pm(k)=\pm\frac{\sqrt{\xi^2(k)+4\Delta\gamma}}{2},\quad |u_\pm\rangle=\begin{pmatrix}
        \frac{-\xi(k)\pm\sqrt{\xi^2(k)+4\Delta\gamma}}{\sqrt{4\gamma^2+(-\xi(k)\pm\sqrt{\xi^2(k)+4\Delta\gamma})^2}}\\
        \frac{2\gamma}{\sqrt{4\gamma^2+(-\xi(k)\pm\sqrt{\xi^2(k)+4\Delta\gamma})^2}}
    \end{pmatrix}.
\end{equation}
Calculate the value of $\eta$
\begin{equation}
    \eta=|u_+\rangle\langle u_+|+|u_-\rangle\langle u_-|=\begin{pmatrix}
        \frac{(\xi^2(k)+2\Delta^2+2\Delta\gamma)}{\xi^2(k)+(\Delta+\gamma)^2}&\frac{\xi(k)(\gamma-\Delta)}{\xi^2(k)+(\Delta+\gamma)^2}\\
        \frac{\xi(k)(\gamma-\Delta)}{\xi^2(k)+(\Delta+\gamma)^2}&\frac{\xi^2(k)+2\gamma^2+2\gamma\Delta}{\xi^2(k)+(\Delta+\gamma)^2}
    \end{pmatrix},
\end{equation}
where satisfies $\eta H\eta^{-1}=H^\dagger$.
The determinant of $\eta$
\begin{equation}
    \det(\eta)=\frac{\xi^2(k)+4\Delta\gamma}{\xi^2(k)+(\Delta+\gamma)^2}.
\end{equation}
Systems with $\eta$-pseudo Hermitian symmetries have real energy spectra for $\det(\eta)>0$
\section{The calculation of response function}
The non-Hermitian is written as
\begin{equation}
    H=H_0+V(t),\quad V(t)=Bf(t),
\end{equation}
where $V(t)$ is perturbation and $H_0$ is non-perturbation Hamiltionian.
There is no restriction here on the Hermiticity of perturbation $B$.

The generalized response function in non-Hermitian system described by  \cite{sticlet2022kubo,pan2020non}
\begin{equation}
    \begin{split}
        \chi_{A,B}(t,t')=&-i\theta(t-t')tr\left\{[A(\tau),B]-\langle A(t)\rangle_0[e^{iH_0^\dagger\tau}e^{-iH_0},B]\frac{\rho_0(t')}{tr{\rho_0(t)}}\right\},
    \end{split}
\end{equation}
where the commutator $[A,B]$ which is called generalized commutator defined by $[A,B]=AB-B^\dagger A$ and $\tau=t-t'$ represents a time interval between initial and final state and the density matrix of unperturbation system Hamiltionian $H_0$ denoted by $\rho_0$ which is time dependent. 


In the non-Hermitian model, the response function has no time transition invariant due to the appearance of the system density matrix $\rho(t)$ at various times, which may have a nonunitary evolution in the absence of the perturbation.

At the initial time, the non-normalized density matrix can be written in terms of the eigenstates $(|\psi_k\rangle, \langle\psi_k|)$ of any Hermitian operator, defined in the Hilbert space of the subsystem, and of their statistical weights $w_k$:
\begin{equation}
    \rho_0=\sum_i w_i|\psi_i\rangle\langle \psi_i|,
\end{equation}
where the $\sum_k w_k=1$.

However if we consider the system is in pseudo-symmetry phase, there is real energy spectrum in the system.
In this case, the density matrix $\rho$ evolves over time in the form
\begin{equation}
    \rho_0(\tau)=\sum_{k}w_k e^{-iH_0\tau}|\psi_i\rangle\langle\psi_i|e^{iH_0^\dagger\tau}=\rho_0.
\end{equation}
So we get the conclusion that density matrix is time independent in the pseudo-symmetry phase. 

With a real eigenspectrum in pseudo Hermitian system, under a time-dependent perturbation $V(t)=B f(t)$, the correlation function can be reduced as
\begin{equation}
    \begin{split}
        \chi_{A,B}(\tau)=&-i\theta(\tau)\left(\langle [A(\tau),B(0)]\rangle_0-\langle A(0)\rangle_0\langle[e^{i H_0^\dagger\tau}e^{-iH_0\tau},B(0)]\rangle_0\right),
    \end{split}
\end{equation}
where the $\chi_{A,B}(\tau)$ is the pseudo Hermitian response of $A,B$. Note that the above formula is in the interaction picture. 
The thermal expectation is defined by $\langle A\rangle_0=tr(\rho_0 A)/tr(\rho_0)$. 
The response function has a value only in the unbroken phase of pseudo-Hermitian symmetry and zero in the pseudo Hermitian symmetric broken phase.
This is due to exact cancellation between the generalized commutator contribution to the response, $[j(k,\tau),j(k,0)]$, and the norm corrections.

For the Kitaev system now, let $j=-e/2\sum_{k}\partial \xi(k)/\partial k \sigma_z$, the response function is written
\begin{equation}\label{response}
    \begin{split}
        \chi(\tau)=&\sum_{k\in PS}i\theta(\tau)\langle[j(k,\tau),j(k,0)]\rangle_0-\langle j(k,0)\rangle_0\langle[e^{iH_0^\dagger\tau}e^{-iH_0\tau},j(k,0)]\rangle_0,
    \end{split}
\end{equation}
where the sum over k can only be calculated in the real spectrum. 
At zero temperature, the system is half full. 

For generic real pairing order parameters $\Delta,\gamma$, both the generalized commutator and the norm correction terms contribute.
We choose the ground state $|u_-\rangle$ of Hamiltionian $H(k)$ in Eq. \eqref{eigen}.
\begin{equation}
    \begin{split}
        \langle[j(k,\tau),j(k,0)]\rangle_0&=\langle j(k,\tau)j(k,0)-j^\dagger(k,0)j(k,\tau)\rangle_0\\
        &=-\frac{8it^2\gamma(\Delta+\gamma)(2\epsilon(k)-\xi(k))\sin(2\epsilon(k)t)\sin^2 k}{\epsilon(k)(2\gamma^2+2\gamma\Delta+\xi(k)(\xi(k)-2\epsilon(k)))},
    \end{split}
\end{equation}
\begin{equation}
    \begin{split}
        \langle j(k,0)\rangle_0=-\frac{2t(2\gamma^2-2\gamma\Delta+\xi(k)(2\epsilon(k)-\xi(k)))\sin k}{2\gamma^2+2\gamma\Delta+\xi(k)(\xi(k)-2\epsilon(k))},
    \end{split}
\end{equation}
\begin{equation}
    \begin{split}
        \langle[e^{iH_0^\dagger\tau}e^{-iH_0\tau},j(k,0)]\rangle_0=\frac{4it\gamma(\gamma-\Delta)(2\epsilon-\xi)\sin(2\epsilon(k)\tau)\sin k}{\epsilon(k)(2\gamma^2+2\gamma\Delta+\xi(k)(\xi(k)-2\epsilon(k)))}.
    \end{split}
\end{equation}
Substituted into Eq. \eqref{response}, the result can be derived:
\begin{equation}
    \chi(k,\tau)=\theta(\tau)\sin^2 k\sin 2\epsilon(k)\tau A[\epsilon(k),\xi(k)],
\end{equation}
\begin{equation}
    \begin{split}
        A[\epsilon(k),\xi(k)]=\frac{128e^2 t^2\gamma^2(2\epsilon(k)-\xi(k))^2}{[\epsilon(k)(2\epsilon(k)-\xi(k))-\gamma(\Delta-\gamma)]^2},
    \end{split}
\end{equation}
where the $\xi(k)=\mu-2t\cos k$ is the kinetic energy term and the $\epsilon(k)=\sqrt{\xi^2(k)/4+\Delta\gamma}$ is the absolute value of energy eigenvalue. 

\section{Viscosity}
The time dependent $\chi(\tau)$ is obtained by summation over all momentum states.
The response function $\chi(\tau)$ of the system exhibits a damped oscillating behavior with a slow decay in the pseudo-symmetry phase of $\gamma>0$.
The $\chi(\tau)$ the response function decays exponentially and vanishes rappidlly in the pseudo-symmetry phase of $\gamma<0$. 
The response in frequency space $\chi(\omega)$ follows by Fourier transforming. 
Putting $\theta(\tau)=\lim_{\eta->0^+}\int\frac{e^{i\omega\tau}}{\omega-i\eta}\frac{d\omega}{2\pi i}$ in Eq. \eqref{response}, we can obtain
\begin{equation}
    \begin{split}
        \chi(k,\omega)&=\int d\tau\chi(k,\tau)e^{i\omega\tau}\\
        &=\frac{1}{2}\sin^2k A[\epsilon(k),\xi(k)]\left(\frac{1}{\omega+2\epsilon(k)+i\eta}-\frac{1}{\omega-2\epsilon(k)+i\eta}\right).
    \end{split}
\end{equation}
Summing over all the real energy spectrum momenta.
\begin{equation}
    \chi(\omega)=\int_{k\in PS}\chi(k,\omega).
\end{equation}
The real part of viscosity is given by the imaginary part of the response function, $\sigma'(\omega)=\chi''(\omega)/\omega$, where the $\chi''(\omega)=[\chi(\omega)-\chi(-\omega)]/2i$ \cite{geier2021non}. 
\begin{equation}
    \begin{split}
        \chi''(\omega)&=-\frac{\pi}{2}\int_{k\in PS}dk \sin^2k A(k)[\delta(\omega-2\epsilon)-\delta(\omega+2\epsilon)]\\
        &=-\frac{\pi}{2}\int_{k\in PS}dk\delta(\omega-2\epsilon(k))\sin^2k A(k)\quad (\omega>0,\epsilon(k)>0)\\
        &=-\frac{\pi}{2}\int_{k\in PS}dk\delta(k-k_i)\frac{\epsilon(k_i)}{t\xi(k_i)\sin k_i} \sin^2kA(k) \quad k_i=k_1,k_2,
    \end{split}
\end{equation}
where the $k_i$ satisfies $\omega=2\epsilon(k_i)$ and the above first line is used $1/(x+i\eta)=P(1/x)-i\pi \delta(x)$. 
\begin{equation}
    \begin{split}
        k_1&=\arccos\frac{\mu+\sqrt{\omega^2-4\Delta\gamma}}{2t},\\
        k_2&=\arccos\frac{\mu-\sqrt{\omega^2-4\Delta\gamma}}{2t},
    \end{split}
\end{equation}
\begin{enumerate}
    \item $\frac{\mu+\sqrt{-4\Delta\gamma}}{2t}<1$ and $\frac{\mu-\sqrt{-4\Delta\gamma}}{2t}>-1$
    \begin{equation}
        \chi''(\omega)=-\frac{\pi}{2}\left(\frac{\epsilon(k_1)}{\xi(k_1)}\sin k_1\frac{128t\gamma^2(2\epsilon(k_1)-\xi(k_1))^2}{[\gamma^2+(2\epsilon(k_1)-\xi(k_1))^2]^2}+\frac{\epsilon(k_2)}{\xi(k_2)}\sin k_2\frac{128t\gamma^2(2\epsilon(k_2)-\xi(k_2))^2}{[\gamma^2+(2\epsilon(k_2)-\xi(k_2))^2]^2}\right).
    \end{equation}
    \item $\frac{\mu+\sqrt{-4\Delta\gamma}}{2t}>1$ and $\frac{\mu-\sqrt{-4\Delta\gamma}}{2t}>-1$
    \begin{equation}
        \chi''(\omega)=-\frac{\pi}{2}\left(\frac{\epsilon(k_2)}{\xi(k_2)}\sin k_2\frac{128t\gamma^2(2\epsilon(k_2)-\xi(k_2))^2}{[\gamma^2+(2\epsilon(k_2)-\xi(k_2))^2]^2}\right).
    \end{equation}
    \item $\frac{\mu+\sqrt{-4\Delta\gamma}}{2t}>1$ and $\frac{\mu-\sqrt{-4\Delta\gamma}}{2t}<-1$
    \begin{equation}
        \chi''(\omega)=0.
    \end{equation}
\end{enumerate}

\section{Phase diagram}
The pseudo-symmetry operator can be written as 
\begin{equation}
    \eta=\begin{pmatrix}
        \frac{(\xi^2(k)+2\Delta^2+2\Delta\gamma)}{\xi^2(k)+(\Delta+\gamma)^2}&\frac{\xi(k)(\gamma-\Delta)}{\xi^2(k)+(\Delta+\gamma)^2}\\
        \frac{\xi(k)(\gamma-\Delta)}{\xi^2(k)+(\Delta+\gamma)^2}&\frac{\xi^2(k)+2\gamma^2+2\gamma\Delta}{\xi^2(k)+(\Delta+\gamma)^2}
    \end{pmatrix},
\end{equation}
\begin{equation}
    \det(\eta)=\frac{\xi^2(k)+4\Delta\gamma}{\xi^2(k)+(\Delta+\gamma)^2}.
\end{equation}
If the system is in pseudo Hermitian symmetric phase, the system has a real energy spectrum.
In the real spectrum, the pseudo Hermitian operator is positive definite, that is $\det (\eta)>0$.
\begin{equation}
    (\mu-2t \cos k)^2>-4\Delta\gamma,
\end{equation}
\begin{equation}
    \begin{split}
        \cos k<\frac{\mu-2\sqrt{-\Delta\gamma}}{2t},\\
        \cos k>\frac{\mu+2\sqrt{-\Delta\gamma}}{2t},
    \end{split}
\end{equation}
For the case where pseudo-Hermitic symmetry is completely destroyed, the system is in the insulator phase. There are conditions:
\begin{equation}
    \begin{split}
        \frac{\mu-2\sqrt{-\Delta\gamma}}{2t}<-1,\\
        \frac{\mu+2\sqrt{-\Delta\gamma}}{2t}>-1,
    \end{split}
\end{equation}
For the case where $\gamma>0$, the low frequency limit viscosity approach to infinite, which means that the system is in the superconducting phase.

\begin{figure}[h]
    \centering
    \includegraphics[width=4in]{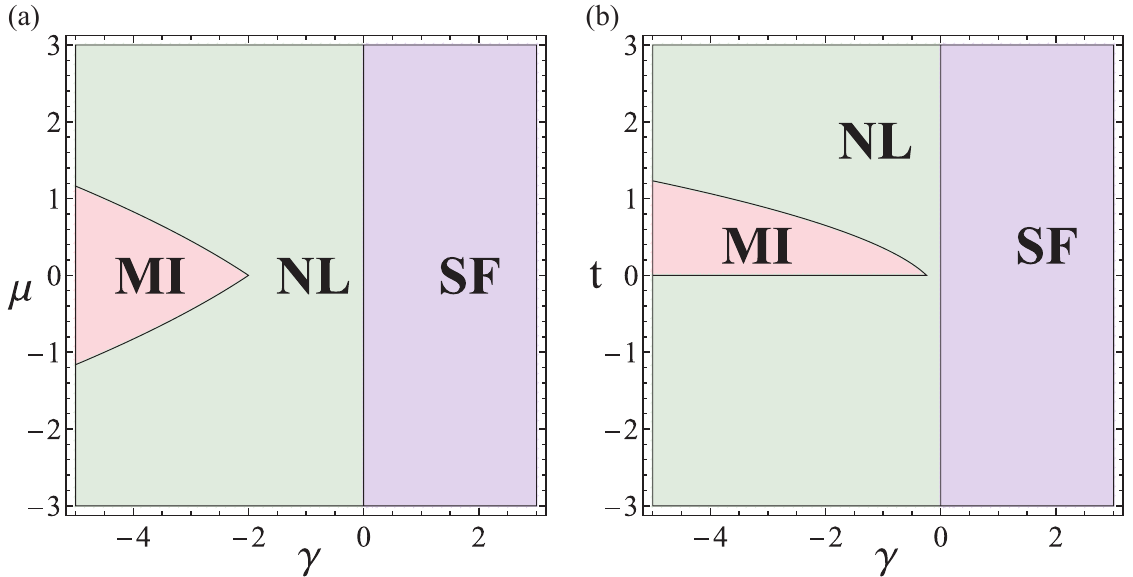}
    \caption{The phase diagram of our system. As $\gamma$ gradually increases, the system changes from an insulator phase to Normal-Liquid(NL) phase and eventually to superfluid(SF) phase. The parameters are set to $t=1,\Delta=0.5$.}
\end{figure}

\end{document}